
\input harvmac

\Title{\vbox{\baselineskip12pt\hbox{Preprint IFUNAM} \hbox{FT96-13}
\hbox{December 1996}\hbox{{\tt hep-ph/9612303n}}}}
{\vbox{\centerline{The q-deformed quark model~: }
	\vskip2pt\centerline{quantum isospin and q-hadrons}}}

\centerline{M.  Klein-Kreisler and M. Ruiz-Altaba\footnote{$^\dagger$}
{(martin@teorica1.ifisicacu.unam.mx, marti@teorica0.ifisicacu.unam.mx)}}
\bigskip\centerline{Instituto de F\'{\i}sica}
\centerline{Universidad Nacional Aut\'onoma de M\'exico}
\centerline{A.P. 20-364, 01000 M\'exico D.F.}

\vskip .3in
We consider the possibility that the  $SU(2)$ isospin symmetry, exact in strong
interactions
 but only approximate in nature, is in fact a quantum group. Using  a doublet
of $q$-quarks, we build the wavefuntions of  $\pi$-mesons, nucleons and
$\Delta$ baryons. We redo the usual quark model computation for the magnetic
moments and mass relations, and everything fits with experimental data. We
find, nevertheless, that it is impossible to parametrize successfully the large
mass difference between the charged and neutral pions with the single   $q$ of
the quantum group $U_q(s\ell(2))$.

\Date{12/96}


\newsec{Introduction}

One of the most  beautiful and old ideas in particle physics is the isospin
invariance of strong interactions, discovered by Heisenberg.  It constitutes
one of the keystones of the quark model, which later became current algebra and
eventually `t~Hooft's standard model of strong interactions, namely quantum
chromodynamics.  Its success in a variety of   predictions makes it a
fundamental conceptual tool for the understanding of many low-energy phenomena
and properties of hadrons. The quantitative predictions, however, are corrected
by the electromagnetic interactions and, of course, the full QCD. It is amusing
and instructive to consider whether a slight deformation of $SU(2)$ can yield
better fits with experiment. Of course, any such deformation  will introduce
one (or more) additional parameters and thus predictivity, along with physical
understanding, will be lost or trivialized unless enough care is exercised. A
beautiful opportunity for deforming the $SU(2)$ of isospin is provided by the
quantum group of Leningrad, Drinfeld and Jimbo, which is the technical context
we shall work in \ref\jim{M. Jimbo, \sl The quantum group $U_q(s\ell(2))$ and
its representations\rm, Springer (1992) Tokyo.}. The basic elements of the
theory of quantum groups we need are briefly reviewed in section~2, along with
some general comments on the possibility of applying quantum groups to
four-dimensional physics. This seems, in principle, non-sense, but, as we shall
argue, we can bypass all objections we could think of in the present context,
with some minor caveats.   Section~3 is devoted to the construction of the
hadron states using the quark  $q$-isospin doublets. Section~4 presents our
computations for the baryons, which go through very smoothly, whereas section~5
is devoted to the pions, where difficulties arise.  It is interesting that,
contrary to naive expectations, the $q$-deformed  $SU(2)$ of isospin cannot
accommodate the large mass difference of pions without invoking explicit
electromagnetic corrections: it is not true that the extra parameter $q$ allows
one to fit anything. Throughout this letter, we restrict ourselves to a world
with only one family (i.e. doublet) of quarks, so we make believe that the
strange and even heavier quarks do not exist or, more soundly, are essentially
completely decoupled  from the static properties of hadrons we investigate. The
final section~6 sums up our conclusions and perspectives.

\newsec{The quantum isospin group and its possible relevance to physics}
Given the $SU(2)$ algebra, it is possible to $q$-deform it to obtain the
quantum group $U_q(s\ell(2))$, with generators $I_+$, $I_-$ and $I_z$, subject
to the commutation relations
\eqn\comm{  \left[I_3, I_\pm \right] = \pm I_\pm \;,\qquad \left[I_+, I_-
\right] = {q^{I_3 } - q^{-I_3} \over q-q^{-1} } }
In the limit $q\to1$, one recovers the usual $SU(2)$ with the help of
L'Hospital's rule.  Let us point out the well-known but crucial fact that the
mere commutation relations \comm\ constitute only, per se, a curious and rather
ugly redefinition of  the generators of  $SU(2)$ \ref\poly{A. Polychronakos,
Mod. Phys. Lett. A5  (1990) 2325 .}. No new physics may arise from just
$q$-deforming (that is, $q$-redefining) the algebra. The meat of the matter
lies in the asymmetric co-products
\eqn\copr{\Delta\left(I_\pm\right) = I_\pm \otimes q^{I_3} + q^{-I_3} \otimes
I_\pm \;,\quad \Delta\left(I_3\right) = I_3\otimes 1 + 1 \otimes I_3}
The choice \comm\ and \copr\ is the canonical one for the quantum deformation
of (the enveloping algebra of) $SU(2)$, and its salient feature is the
asymmetry in the coproducts for $I_\pm$, which disappears in the classical
$q\to1$ limit.  It is possible to redefine the generators such that the algebra
\comm\ is the classical one, but then the coproducts \copr\ become complicated
and remain {\sl asymmetric}. Inversely, the usual (classical) $SU(2)$ can be
defined by \comm, but then the coproduct is complicated and {\sl symmetric}.

The point we wish to emphasize is that we want to use the full algebraic
structure of the quantum group (along with its   other  unmentioned features),
not only the awkward \comm. By this we mean that we really want to imagine for
a moment that the action of, say, the isospin lowering generator $I_-$ on a
two-particle state distinguishes between the first and the second particles:
indeed, since its action on the state is precisely $\Delta \left(I_-\right)$,
and $\Delta $ is asymmetric, we must have a way of ordering the two particles
so that the results we get are physically meaningful.  In general, this seems
impossible if the particles are asymptotic states. Even for them, the
consistent use of  a clever Drinfeld twist  allows the preservation of  the CPT
theorem with only  Fermi or Bose statistics  surviving \ref\germanicos{G. Fiore
and P. Schupp , Nucl. Phys. B470 (1996) 211.} but the problem of interpretation
remains.  For our application of the quantum group to strong interactions, we
shall limit ourselves, in this letter, to single particle properties
\ref\HRsingle{J.Wess, O. Ogievetskii, M. Pillin, W.B. Schimdke and B. Zumino,
Proceedings of the 22nd Intl. Conf. on Differential Geometry Methods in
Theoretical Physics, World Scientific  (1993) Singapore; \semi H. Ruegg and
V.N. Tolstoi, Lett. Math. Phys. 32 (1994) 85.} such as the mass and the
magnetic moment.  We do not solve the problem but bypass it.

Still, the same problem of ordering crops up again when we start talking about
the quarks inside the hadron. We will really think that in a $\pi^+$ it makes
sense to speak of the $u$ quark as being before or after the $\bar d$ quark.
The two alternatives are not completely unrelated: the passage from one to the
other  is just a similarity transformation, essentially  a messy braiding with
the famous $R$-matrix. Since quarks are confined, we feel justified in applying
to them funny prescriptions. This practice mimicks closely the  symmetrizations
one does in the standard quark model  to get the relevant wavefunctions for
hadrons.  Of course, the kets will be related to the conjugates of the bras
with the extra braiding factors of the $R$-matrix, but after the party is over
we still end up with  expressions for the hadrons in terms of the quark states
and the deformation parameter $q$. Whether $u$   or $\bar d $ comes first
amounts to exchanging the coproduct $\Delta$ with its transpose $\Delta'$ (one
is the conjugate to the other through the $\cal R$-matrix)  or,  equivalently,
to interchanging $q$ with $q^{-1}$.  In any given analysis, we stick to a
choice of conventions and keep in mind that the same computation by someone
else with the same result might need the redefinition $q\leftrightarrow
q^{-1}$.  To illustrate the point with a touch of chutzpah, we shall use
$\Delta'$ for mesons and $\Delta $ for baryons below.

The solution to the ordering problem for quarks inside a hadron is physically
simple: we go to the infinite-momentum frame and work there. This agrees, of
course, with the usual quark model framework \ref\quarkmod{F. Halzen and A.D.
Martin, {\sl Quarks and leptons: an introductory course}, Wiley (1984) New
York.}. What is new, is that  we add an ordering to the quarks in this
infinitely boosted frame.
This trick has been used  in field theory to derive an integrable
two-dimensional (XXZ) model  from QCD \ref\Lipatov{ L.N. Lipatov, JETP Lett. 59
(1994) 596\semi L.D. Faddeev, Phys. Lett. 342B (1995) 311.}.

\newsec{$q$-hadrons from $q$-quarks}

The doublet of quarks  $\pmatrix{|u>\cr |d>\cr}$ sits in a fundamental irrep of
$U_q(s\ell(2))$, so that $I_-|u>=|d>$ and so on. The doublet of antiquarks is
$\pmatrix{|{\bar d}>\cr -|{\bar u}>\cr}$
where the $-$ is conventional and allows one to read off the $G$-parity of pion
states effortlessly (nothing of what we shall do depends on this sign).  Taking
the tensor product of these two doublets yields a $q$-isovector and a
$q$-isosinglet. The highest weight in the $I=1$ $q$-triplet,  is $|\pi^+> = |u
>\otimes|{\bar d}> = |u {\bar d}> $. We choose the  ordering convention
quark-antiquark.  Also, we choose $\Delta'$ as coproduct, so that the $I_3=0$
partner of  $|\pi^+>$ is proportional to
\eqn\pip{ \Delta'  \left( S_- \right)  |u >\otimes|{\bar d}> = -q^{1/2}  |u
>\otimes|{\bar u}> +
q^{-1/2}  |d >\otimes|{\bar d}> }
The normalization of this state requires the use of a conjugation, which
reverses the order of the states in the tensor product, and thus involves the
$R$-matrix.  To avoid pathologies, the value of $q$ (not as a ``symbol'') must
be either real or a  pure phase. The result is, predictably,
\eqn\pizero{  |\pi^0> =  {1\over \sqrt{q+q^{-1} }} \left( -q^{1/2}  |u  {\bar
u}> +
q^{-1/2}  |d  {\bar d}>  \right) }
And, of course, $|\pi^->  = |d {\bar u}> $.

The $\Delta $ spin-3/2 resonances (with coproduct $\Delta $ as advertised) are
similarly obtained by acting on  $|\Delta^{++}> =|uuu>$ with $(1\otimes
\Delta)\Delta(I_-) = (\Delta\otimes 1) \Delta(I_-)$\foot{Do not confuse the
baryons $\Delta^{++}$, $\Delta^+$, etc.  with the coproducts $\Delta$ and
$\Delta' =P\circ \Delta$!}. We find
\eqn\deltas{  \eqalign{ &|\Delta^+>  = {1\over \sqrt{q^2 + 1 + q^{-2}}} \left(q
|duu> + |udu> + q^{-1} |uud>\right) \cr
&|\Delta^0>  = {1\over \sqrt{q^2 + 1 + q^{-2}}} \left(q^{-1} |udd> + |dud> +
q^{-1} |ddu>\right) \cr}}
and $|\Delta_->=|ddd>$.

The nucleon states require some attention. In the decomposition $1/2 \otimes
1/2 \otimes 1/2= 1/2 \oplus 1/2 \oplus 3/2$ we wish to identify the nucleon
doublet as  a  state symmetric in the two quarks with  spin $\Uparrow$.
Although we have not written out the spin part of the wave-function, for the
mesons it was something like $\Uparrow\Downarrow$ (along with the appropriate
symmetrization) and for the $\Delta's$ it is $\Uparrow\Uparrow\Uparrow$. Now,
if we look at the $\Uparrow\Uparrow\Downarrow$ piece of the nucleon
wave-function, we must require that the flavor counterpart be $q$-symmetric in
the first two quarks (which are symmetric in the spin wavefunction). Using the
fact that $\Delta(I_-)|uu>=q^{-1/2}|ud> +q^{1/2} |du>$, we require the nucleon
wave-functions to be
\eqn\nucl{ \eqalign{ &|p> \sim  q^{-1/2}|udu> +q^{1/2} |duu> + P |uud> \cr
&|n> \sim  q^{-1/2}|udd> +q^{1/2} |dud> + N |ddu> \cr} }
with $P$ and $N$ to be determined from the doublet condition.  After
conjugating with the $R$-matrix and normalizing, the result is again simply
\eqn\proton{ \eqalign{ &|p> ={1 \over \sqrt{ 1+q^{-2} + \left(1+q^2 \right)^2 }
} \left[  q^{-1}|udu> +  |duu> - \left(1+q^2 \right)  |uud> \right] \cr
&|n> ={1 \over \sqrt{ 1+q^{2} + \left(1+q^{-2} \right)^2 } } \left[  q |dud> +
|udd> - \left(1+q^{-2} \right)  |ddu> \right] \cr } }
To end this section, note that the above wavefunctions are valid for $q$ real.
Similar expressions hold for $q$ a pure phase (just the normalizations change
slightly). In what follows, we shall always quote the results for $q$ real
only.

\newsec{$q$-baryons}
We are now equipped to compute  masses and magnetic moments.  The mass operator
$M$ and the charge operator $Q$ are postulated to have trivial coproducts (like
$I_3$), and to be diagonal in the constituent quarks. Of course, the valence
quark masses we use are different in each isospin multiplet, so for instance
the value of $m_u$ from nucleons need not be the same as that from $\Delta$'s.
We note
\eqn\massdelta{ m_{\Delta^{++}} =  <\Delta^{++}| M\otimes 1\otimes 1 + 1\otimes
M\otimes 1 + 1\otimes 1\otimes M |\Delta^{++}> } and similarly for the other
states, and find that the $q$ dependence drops out:
\eqn\masadel{ m_{\Delta^{++}} =3m_u \;,\quad
m_{\Delta^{+}} =2m_u+m_d \;,\quad
m_{\Delta^{0}} =m_u+2m_d \;,\quad
m_{\Delta^{-}} =3m_d }
so that  the prediction from $q$-isospin is the same as the usual one in the
traditional quark model, namely
\eqn\hhh{  m_{\Delta^{++}} - m_{\Delta^{+}}  =
 m_{\Delta^{+}} - m_{\Delta^{0}}  =
 m_{\Delta^{0}} - m_{\Delta^{-}}  = m_u-m_d \equiv - \delta}
Although the actual values of  the parameters $m_u$ and $m_d$ are of no
interest,  their difference $\delta =m_d - m_u$ should be more or less
glue-free and may be identified with the mass difference obtained from nucleons
(again the $q$'s disappear):
\eqn\massnuc{ m_p = 2m_u +m_d \;,\quad  m_n = m_u + 2m_d}
Thus, \eqn\dddee{\delta = m_n-m_p =  \, {\rm 1.29 MeV} }
which agrees with the spread in masses of the $\Delta $-resonances\ref\PDG{R.M.
Barnett et al., Particle Data Book,  Phys. Rev. D54  (1996) 1 .}, as it should
since the quark model works.

To compute the nucleon magnetic moments, we look at the piece of the
wavefunction with spins aligned according to $\Uparrow\Uparrow \Downarrow$, so
that
\eqn\mup{ \mu_p = <p| \mu \otimes 1  \otimes 1 + 1\otimes \mu \otimes 1 -
1\otimes 1 \otimes \mu |p> } and similarly for $\mu_n$. We find the following
gruesome expressions
\eqn\muuuu{ \eqalign{
& \mu_p = {1\over 1+ q^{-2} + (1+q^2 )^2 } \left\{ 2(1+q^2)^2 \mu_u - \left[
(1+q^2)^2 - (1+q^{-2}) \right] \mu_d \right\} \cr
& \mu_n = {1\over 1+ q^{2} + (1+q^{-2} )^2 } \left\{ 2(1+q^{-2})^2 \mu_d -
\left[ (1+q^{-2})^2 - (1+q^{2}) \right] \mu_u \right\} \cr }}
Therefore, using now the point-particle expression for the magnetic moment of a
quark, $\mu=Q/m$, we find the interesting
\eqn\mumu{ {\mu_n\over\mu_p} = -2 \; {  (m_u +m_d) (1+q^{-2}- q^2 ) + q^2 m_u
\over   (m_u +4m_d) (1-q^{-2}+ q^2 ) + 4q^{-2} m_d } }
Experimentally, ${\mu_n\over \mu_p} =-.684979$ and the quark model prediction
is $-2/3$. Under the (wrong) hypothesis that $m_u=m_d$, the $q$-quark model
fits the datum
with $q=.992$. As emphasized repeatedly, this value of $q$ is physically
equivalent to the one obtained from $\Delta'$, namely $q=1.008$. Using the mass
difference $\delta$ noted above \dddee, we get instead $q=0.991$.

\newsec{$q$-pions}
In contradistinction with the baryon case,  the $q$'s in the meson
wave-functions may show up in their mass formulae, but only if $q$ is a phase.
For real $q$, from the wavefunction \pizero, we find \eqn\mmmm{ m_{\pi^\pm} =
m_{\pi^0} = m_u + m_d}
Meaning that in the $q$-deformed quark model, the neutral and charged pions
remain degenerate. But for $q$ a pure phase, we find
\eqn\kkn{ m_{\pi^\pm} = m_u +m_d \;,\qquad m_{\pi^0} = {2\over q  + q^{-1}}
\left( q^{-1} |d {\bar d}> + q |u{\bar u}> \right) }
where we have used, as advertised,  the coproduct $\Delta'$.  Note that the
prefactor is the inverse of cos$\alpha$, if $q={\rm e}^{ {\rm i} \alpha}$, so
it  tends to make the neutral pion heavier than the charged ones. Also, the
mass turns out to be complex unless $m_u=m_d$,  in which  case all the
$q$-dependence cancels out and we are left with the degenerate case again.

To solve this conundrum, one must  include electromagnetic corrections. Very
simply, we take the additional contribution to the mass formulae to be given by
the electrostatic potential between the two quarks in the pion, as if they were
hanging out at some distance $R$ from each other.  It turns out that
\eqn\deded{\delta_E m_{\pi^\pm} = 2 E_\pm \;, \qquad \delta_E m_{\pi^0 } =
-\,{4q + q^{-1} \over q+ q^{-1}} E_0}
where $E_x = {4\pi \alpha \over 9} {1\over R_x}$ and we  would expect
$R^{-1}_x\sim m_{\pi^x}$.
Taking the experimental value of $\Delta m_\pi = m_\pi^\pm - m_\pi^0 =$ 4.59
MeV, takes us then to the real value $q= 0.324  $.

The reader  ought to be relieved that this numerology is over. The clear
conclusion is that the low mass of the neutral pion cannot be naturally
accommodated in the simple $q$-deformed isospin symmetry.   Trying to explain
why (or rather, parametrize how) the $\pi^0$ is lighter than the $\pi^\pm$ was
the original motivation for this work, and it seems to be impossible: the value
of $q$ derived from the nucleons' magnetic moments does not account for it.

\newsec{Conclusions and outlook}
As noted in section~2 above, we have restricted ourselves to static predictions
of the ($q$-deformed) quark model.  They are   satisfactory for the nucleons
but not so for the pions.  We can fit the parameter $q$ from the magnetic
moments of the nucleons, and recover their masses in the usual way.  It may be
that the $\pi^0$ is not amenable to such a simple picture, exceptionally rather
than generically. After all, it is the lightest hadron, an almost massless
goldstone boson,  with the extremely long lifetime involving the chiral
anomaly, and so on. Let us emphasize  that $q$ disappears from the mass
formulae and can be brought in only through the very \sl ad hoc \rm
electrostatic corrections, which work  more or less in the usual quark model
anyway. The   extension to dynamics of $q$-isospin involves, as a first step,
how to include the parameter $q$ in the amplitudes for $\pi-N$ scattering:
there $q$ would just work  as an additional parameter to better the fits
between the quark model and data, but the conceptual difficulty of
$q$-deforming  hadronic interactions seems rather formidable (even in the
light-cone).

It  would be very nice if quantum groups provided an explanation for the
celebrated $I=1/2$ rule, but we have no inkling on how this may come about
(back-of-the-envelope estimates with $q$ a root of unity do not work).  The
traditional  wordplay ``spin from isospin'' is also food for thought, or sleep
in the present context.   We believe, however,  that  the test of  $q$-isospin
will come from its extension to a $q$-deformed flavor $SU(3)$, where the large
mass difference between the strange quark and the light $q$-isospin doublet
allows for much cleaner computations ($q$ should be rather different from one)
and the pion mass difference is of no great import.  At the very least, one
should be able to come up with a $q$-deformed Gell-Mann--Okubo mass formula.

We believe that, despite the relative meagerness of our quantitative results,
it is worthwhile to explore where, in four dimensions, quantum symmetries may
show up. The more hidden the symmetry, the better, and thus quarks seem obvious
candidates, for these explorations.

This work is partially supported by CONACYT.

\listrefs
\bye